# Loss-driven miniaturized bound state in continuum biosensing system

Jiacheng Sun[1,2,3]†, Fajun Li[5]†, Xudong Wang[4], Jing He[2,3], Dangwu Ni[2,3], Lang Wang[2,3], Shaowei Lin[6], Min Qiu[3,4], Jinfeng Zhu[5]*, Liaoyong Wen[2,3,4]*.

[1]College of Information Science and Electronic Engineering, Zhejiang University, Hangzhou, 310027, China.

[2]Research Center for Industries of the Future (RCIF), School of Engineering, Westlake University, Hangzhou 310024, China.

[3]Zhejiang Key Laboratory of 3D Micro/Nano Fabrication and Characterization, School of Engineering, Westlake University, Hangzhou, Zhejiang 310030, China

[4]Westlake Institute for Optoelectronics, Fuyang, Hangzhou, Zhejiang 311421, China.

[5]Institute of Electromagnetics and Acoustics and Key Laboratory of Electromagnetic Wave Science and Detection Technology, Xiamen University, Xiamen, 361005, China

[6]Department of Nuclear Medicine, the First Affiliated Hospital of Xiamen University, School of Medicine, Xiamen University, Xiamen, 361003, China

†These authors contributed equally.

*Corresponding author. E-mail: jfzhu@xmu.edu.cn; wenliaoyong@westlake.edu.cn

**Abstract:**
Optical metasurface has brought a revolution in label-free molecular sensing, attracting extensive attention. Currently, such sensing approaches are being designed to respond to peak wavelengths with a higher $Q$ factor in the visible and near-infrared regions. Nevertheless, a higher $Q$ factor that enhances light confinement will inevitably deteriorate the wavelength sensitivity and complicate the sensing system. We propose a $Q$-switched sensing mechanism, which enables the real part of the refractive index to effectively perturbate the damping loss of the oscillator, resulting in a boost of peak intensity. Consequently, a higher $Q$ factor in $Q$-switched sensor can further enhance the peak sensitivity while remaining compatible with broadband light sources, simultaneously meeting the requirements of high performance and a compact system. This is achieved in a unique 3D bound-state-in-continuum (BIC) metasurface which can be mass-produced by wafer-scale aluminum-nanoimprinting technology and provides a peak intensity sensitivity up to 928 %/RIU. Therefore, a miniaturized BIC biosensing system is realized, with a limit of detection to $10^{-5}$ refractive index units and 129 aM extracellular vesicles in clinical lung cancer diagnosis, both of which are magnitudes lower than those of current state-of-the-art biosensors. It further demonstrates significant potential for home cancer self-testing equipment for post-operative follow-up. This $Q$-switched sensing mechanism offers a new perspective for the commercialization of advanced and practical BIC optical biosensing systems in real-setting scenarios.

# Main

Label-free optical biosensors are at the forefront of modern and future disease management, offering rapid point-of-care (POC) diagnosis and continuous monitoring of biomarker or therapeutic drug levels. (*1, 2*) Among the most prominent commercial devices in this category is the surface plasmon resonance (SPR) biosensor. However, SPR biosensors require oblique incidence to excite surface plasmons, necessitating a bulky prism and angle alignment system, thereby significantly increasing the system's footprint and complexity (*3*). To overcome these limitations, nanophotonic metasurfaces have emerged as an alternative sensing module, enabling the downsizing of optical biosensors through the use of normal incidence (*4, 5*). The basic sensing mechanism of nanophotonic biosensors can be summarized as a phase diagram in **Fig.1a**. Refractometric affinity biosensor usually applied in visible and near-infrared spectroscopy can be classified in 1$^{st}$ quadrant which links the response of real part of frequency change ($\Delta w$) to real part of refractive index change ($\Delta n$). For surface-enhanced infrared spectroscopy in 3$^{rd}$ quadrant, molecules can be detected directly by the signal peak intensity due to their inherent energy absorption. This is the imaginary part of frequency change ($\Delta \gamma$) induced by imaginary part of refractive index change ($\Delta k$). Currently, for refractometric affinity biosensor, researchers have extensively sought resonance properties with high-$Q$ factors for better sensing performance through various configurations, such as surface lattice resonances (SLR) (*6–9*), gap surface plasmons (*10–12*), Fano-like modes (*13*), Mie-type resonances (*14–16*), bound states in the continuum (BIC) (*17–24*).

However, the sensing mechanism of these high-$Q$ biosensors is still trapped in the 1$^{st}$ quadrant as a single oscillator model (**Fig.1b**), while the content in 2$^{nd}$ and 4$^{th}$ is empty. Consequently, pursuing narrower signal bandwidth with tightly confined optical mode and energy radiation will inevitably degenerate the frequency sensitivity and signal modulation depth (**Fig.1c**), imposing greater demands on the illumination and transducer setups to measure those high but constant-$Q$ factor (**Fig.1d**). This conflicts with efforts toward miniaturization and practicality of biosensing system due to the following reasons: Firstly, the high-$Q$ resonance peak necessitates coherent light source and spectrometer with high resolution to distinguish subtle resonance frequency shifts, complexing the sensing system (*17*). Secondly, although spectrometer-free biosensors have been developed, they still rely on bulky and high-cost tunable narrowband light sources, such as tunable lasers or filters, for illumination (*25, 26*). Because accurate frequency alignment is required for high-$Q$ resonance. With dynamic frequency during sensing, it easily misses alignment with the static illumination frequency, leading to a frequency mismatch. In contrast, more compact broadband light sources are incompatible with high-$Q$ resonances due to insufficient signal overlap resulting from bandwidth mismatch (*27, 28*). Additionally, high-$Q$ resonance like quasi-BIC in asymmetric structure is highly sensitive to the nanostructure's geometry (*29*), which requires highly accurate fabrication method. This further restrains the practicability of high $Q$ biosensing system in real applications which require low-cost and compact set-ups with massive production of sensor chips such as clinical cancer diagram.

We propose a $Q$-switched sensing mechanism which fills the gap of 4$^{th}$ quadrant in sensing phase diagram. It is realized by two strongly-coupled oscillators which link the $\Delta n$ to $\Delta \gamma$, facilitating a loss-driven response to the refractive index perturbation (**Fig.1e**). In this case, a high-$Q$ resonance can offer a broadband response to $\Delta n$ (**Fig.1f**) with rapidly switched radiative $Q$ factor ($Q_r$) from high to low, crossing the nonradiative $Q$ factor ($Q_n$) as shown in **Fig.1g**. During this process, a significant peak intensity response can be boosted. This mechanism perfectly solves the bandwidth and frequency mismatches, and facilitates LED-compatible miniaturized BIC biosensing system with limit of detection (LOD) as low as $10^{-5}$ refractive index units and 129 aM extracellular vesicles in clinical lung cancer diagnosis, revealing advanced sensing performance (**Fig.1h**). Furthermore, the 3D-BIC metasurfaces can be produced on 8-inch wafers, using a cost-effective aluminum 3D imprinting with stripping techniques (**Fig.1i**). Therefore, our device shows significant potential as a miniaturized and low-cost home self-testing equipment, like cancer postoperative follow-up (**Fig.1j**). $Q$-switched sensing mechanism provides a new perspective for developing high performance and compact optical biosensing systems in real applications, promoting the commercialization of BIC metasurface biosensor.

**Theoretical model of $Q$-switched sensing mechanism**

The eigenmodes of two oscillators in a strong coupling system are typically described by a 2×2 Hamiltonian matrix (*30*):

$$H = \begin{pmatrix} \omega_1 + i\gamma_1 & g \\ g & \omega_2 + i\gamma_2 \end{pmatrix} \quad (1)$$

where $\gamma_1$ and $\gamma_2$ represent the damping rates of the oscillators, $\omega_1$ and $\omega_2$ represent the frequncies of the oscillators and $g$ is their coupling strength. When $\omega_1=\omega_2$, the Rabi splitting is given by:

$$\Omega_R = 2\sqrt{g^2 - (\gamma_1-\gamma_2)^2/4} \quad (2)$$

Here, ambient refractive index *Δn* is introduced to turn the system from zero to non-zero detuning (*Δw*). In this case, $\gamma_1$ and $\gamma_2$ have to offer corresponding feedback (*Δγ*) to compensate such *Δn* induced detuning (**Fig.1k**). Therefore, we can link the *Δn* to *Δγ* with proper coefficients in strong coupling system. This relationship can be further translated to the $Q$ factor ($Q = w/2\gamma$), thereby realizing the refractometric $Q$-switched equation:

$$Q_r = \frac{\omega}{2(\gamma_2' - \left(\sqrt{4g^2 - (\Omega_R + \Delta n (S_1 - S_2))^2} - \sqrt{4g^2 - \Omega_R^2} - i(\omega_1 - \omega_2)\right) - \gamma_n)} \quad (3)$$

where $S_1$ and $S_2$ represent the frequency sensitivities of the oscillators 1 and 2, $\gamma_2'$ is the original damping rates of oscillator 2 at zero detuning, $\gamma_n$ is determined by the intrinsic material property (detailed derivation is provided in **Method** and **Fig. S1-3**). In order to make the analytic solution more realistic, we deduced the corrected refractometric $Q$-switched equation (**Method** and **Supplementary equation E18**) to calculate the analytical results in **Fig.1l**. And the peak intensity change (*ΔI*) induced by $Q$-switched process can be calculated according to one-port system absorption equation:

$$Abs = \frac{2Q_r Q_n}{(\omega - \omega_r)^2 + (Q_r + Q_n)^2} \quad (4)$$

As the critical coupling happens at $Q_r=Q_n$, where the absorption of system maximized, the approaching of $Q_r$ to $Q_n$ will induce significant increasing of peak intensity. Therefore, when the $Q_r$ is switched by *Δn* from high to low, crossing the $Q_n$, the *ΔI* responses intensively with subtle *Δn* and then meets maximum (**Fig.1m**). Besides, with the constrain of initial radiative damping rate $\gamma_{r0}$, imitating the introduction of BIC enhanced $Q_r$, the *ΔI* will become more sensitive to *Δn*, which indicates a BIC-enhanced sensing performance. The theoretical optimal peak intensity sensitivity can reach to > $10^3$ %/RIU.

The above process is named as *Q*-switched sensing and such boosting of peak intensity change can be applied for ultrasensitive biomolecular sensing. Furthermore, it is a general sensing model, not only applicable in strong coupling system (**Fig.S4**), which offers a solution for high *Q* resonance to be compatible with broadband light sources, promoting the partibility of high *Q* biosensors.

**Design and fabrication of 3D-BIC metasurface**

In order to satisfy the condition for *Q*-switched sensing in strong coupling system, a specific nanostructure is needed to be designed with two strongly coupled modes with markedly different responses ($\Delta w_1 \gg \Delta w_2$, $\Delta \gamma_1 \ll \Delta \gamma_2$) to $\Delta n$. This means that one mode should still behave like a conventional affinity sensor with $\Delta w > 0$, while the other mode should be insensitive to the $\Delta n$ with $\Delta w \approx 0$. Our 3D-BIC metasurface can perfectly match these conditions. This 3D-BIC metasurface comprises two sets of longitudinally displaced Au nanoparticle arrays embedded within a transparent resin substrate (NOA83H, n = 1.56), along with an Au nanohole array attached to the surface. The system is illuminated from the substrate side, operating as a reflective biosensor. This design effectively segregates the light path from the microfluidic system, mitigating interference from analytes and fluid disturbances. Consequently, it ensures a more stable signal in various environments, including absorptive or turbid conditions (*31*). The calculated mode profile in **Fig.1n** has sufficient mode overlap with nanoparticles in the superstrate, offering $\Delta w$. Notably, another mode in **Fig.1o** mainly confined inside the substrate with $\Delta w \approx 0$, but offering significant $\Delta Q_r$.

The reason for designing a 3D spatial-asymmetric configuration instead of using common geometry-asymmetry quasi-BIC (qBIC) structure is to avoid mode crosstalk. As shown in **Fig.2a**, because geometry change will significantly influence the resonance frequency, applying commonly used asymmetric geometry in such multi-mode system will cause crosstalk and disturb the qBIC manipulation (*32-34*). On the contrary, instead of changing the geometry, directly shifting the spatial location of nanoparticles in vertical can successfully lead to the generation of qBICs at upper branch (qBIC$_U$) and lower branch (qBIC$_L$) with frequency-robustness and polarization-insensitivity (**Fig.2b**). Detailed mode analysis is provided in **Fig. S8-10**.

The 3D spatial-asymmetric BIC metasurface seems challenging to produce using conventional fabrication techniques such as electron beam lithography and focused-ion-beam lithography. However, it can be readily fabricated on silicon substrates using our binary-pore anodic aluminum oxide (BP-AAO) template, which allows for precise control of out-of-plane displacement ($\Delta Z$) (*35–37*). Notably, scalable 3D-BIC metasurface chips can be manufactured on aluminum wafers through a combination of aluminum-3D imprinting and direct stripping. Representative 6- and 8-inch aluminum wafers, along with their stripped 3D-BIC metasurface chips, are shown in **Fig. 1i**. Scanning electron microscope (SEM) and atomic force microscopy (AFM) images in **Fig. 2c** and **d** confirm the successful fabrication of two (A and B) sets of Au nanoparticle arrays with a longitudinal displacement of 140 nm. A more detailed fabrication process and characterization are provided in **Fig. S5-7**.

Microscopy optical spectrum measurement (**Fig. S11**) measured a typical qBIC behavior for dual resonances, with the linewidth transforming from undetectable to conspicuous as $\Delta Z$ increases from 0 to 140 nm (**Fig. 2e**). This measured trend closely aligns with our calculated results. Notably, the calculated results incorporate a 1.2-fold imaginary part of the Au refractive index, $Im(n_{Au})$, to imitate the introduction of material loss from real fabrication. Nevertheless, the raised $Im(n_{Au})$ did not change the frequency of qBIC$_2$ obviously, demonstrating its high robustness to material loss (**Fig. S13**).

The frequency and mode evolution of two energy branches are shown in **Fig.2f** and **g**. By increasing the thickness of TiO$_2$ layer upon the Au film, the electric field distribution of UP will transfer from superstrate to substrate, while the electric field distribution of LP acts oppositely. This reveals the hybrid of two SPP modes in this strong coupling system. With the introduction of 3D-BIC configuration ($\Delta Z$=100), a larger Rabi splitting is observed and a FW-BIC appears, which confirms the effective constraining of radiative loss (see detailed analysis in **Fig.S14**) (*38-40*).

Simulated $Q_r$ and peak intensity change of qBIC$_U$ (**Fig.2h**, **i** and **Fig.S15-16**) verify the analytical results in **Fig.1l** and **m** that with the participation of BIC manipulation, raised $Q_r$ can bring larger intensity response for optimizing the $Q$-switched sensing performance. And this is further confirmed in experiments, by depositing TiO$_2$ and measuring the spectrum cyclically. As shown in **Fig.2j, k** and **Fig.S17**, larger Rabi splitting and intensity response range in experiment confirms the BIC-enhanced $Q$-switched sensing performance, which offers an efficient tool for optimization.

**$Q$-switched sensing performance and system miniaturization**

In order to evaluate the bulk sensing performance of $Q$-switched sensor in short-wavelength infrared (SWIR), the refractive index of superstrate is perturbated (**Fig.3a**). With small detuning between two branches, qBIC$_U$ exhibits significant intensity growing. Then, the effectiveness of $Q$-switched sensing across VIS, NIR, and SWIR is further proved both in simulation and experiment (**Fig.S20** and **Fig.3b**). The corresponding experimental $Q_r$ and peak intensity change are extracted in **Fig.3d** and **e**. Notably, larger $Q$ factor will induce better peak intensity sensitivity which could reach to a linear 928 %/RIU in experiment ($S_I = \Delta I/\Delta I_0/\Delta n$). This is counterintuitive compared with spectral-shift sensor whose peak wavelength sensitivity will be inevitably degenerated with higher $Q$ factor (**Fig.3f**). Therefore, $Q$-switched sensing reveals a way for optimizing peak sensitivity by increasing the $Q$ factor with more confined light field. Such strategy could be further expanded to enhance the sensing performance in other coupled system, like exceptional points sensors in non-Hermitian systems (*42,43*).

The compatibility of $Q$-switched sensor with various light sources is further demonstrated in a spectrometer-free image system. Using a lithographic overlay method, SiO$_2$ strips with thickness gradients from 10 to 30 nm were fabricated (**Fig.3g** and **Fig. S21**). Illuminated by a tunable 2 nm bandwidth light source in water, the reconstructed morphology in **Fig.3h** with varying center wavelength tuned from 628 to 640 nm (**Fig. S22**). This reveals that for capturing narrow band signals, complicated tunable light sources are necessitated. However, broadband illuminations (10 and 25 nm) at 640 nm effectively reconstruct the gradient morphology, with clear intensity differentiation (**Fig. 3i** and **j**) and outstanding large-scale uniformity (**Fig. S21a** and **b**). Additionally, the system achieves a limited surface image resolution of SiO$_2$ to 3.5 nm (vertical resolution) (**Fig. S23**). Thus, the $Q$-switched sensing mechanism can facilitate a simplified high-performance system integrating high-$Q$ metasurface with a broadband light source. This approach eliminates the illumination requirement for expensive, bulky tunable filters or lasers.

Benefitted by the compatibility with broadband light source, a LED-driven miniaturized BIC biosensing system is achieved (**Fig.1j**). It translates the optical intensity change to electric signal, which can monitor the environment refractive index change in real-time (**Fig.3k**). The miniaturized system achieves a LOD 5.1×10$^{-5}$ for bulk refractive index changes (**Fig.3l**), even surpassing most advanced high $Q$ biosensor in spectrometer measurement and competitive to current commercialized products (*25,44*). The kinetic curves of interaction between IgG and ProA in **Fig.3m** shows highly stable responses from our miniaturized BIC biosensing system with equilibrium dissociation constant, $K_D$, as low as 3.55 nM. Therefore, it maintains great potential for real-setting scenarios out of the laboratory, such as point-of-care or home self-test equipment.

**Clinical lung cancer diagnosis with DNN assistance**

To emphasize the clinical potential of miniaturized BIC biosensing system, it is applied for lung cancer screening. Lung cancer (LC), a leading cause of cancer-related deaths, has tumor-derived exosomes as promising biomarkers for liquid biopsy (*45, 46*). **Fig.4a** and **b** present the clinical profiling of our 3D BIC metasurface for lung cancer diagnosis via LC-derived exosomes. The peak intensity of $Q$-switched resonance exhibits a conspicuous response to each bio-functional step (**Fig.4c**). Notably, it achieves a limit of detection (LOD) of 129 aM for LC-derived exosome sensing, significantly surpassing current state-of-the-art biosensors (**Table S1**) (*47–57*).

Spectral signal response variations in **Fig.4d** are measured from 25 lung cancer patients and 15 healthy controls (see **Method**), showing that no matter using peak searching or integration data process method, their overall performance remains consistent. A dichotomous approach with a defined decision boundary can achieve an 85% accuracy with a high area under the curve (AUC) (**Fig. S30**). However, the presence of numerous suspected cases near the decision boundary affects the practical reliability of this method for clinical diagnosis.

To address the challenge of signal randomness and binary decision-making in cancer diagnosis, we utilized a DNN model to enhance assessment accuracy. A series of $\Delta I$ spectral data (**Fig. 4e**) collected at three different bandwidths (2, 5, and 10 nm) are trained as three individual models for predicting cancer risk (**Fig. 4f**). Details of the DNN model are available in **Method** and **Fig. S31, 32**, where the generalizability and robustness of DNN models are trained. Fluctuations in test accuracy were attributed to certain intractable cases leading to false positives or false negatives (FP/FN) as illustrated in the DNN-predicted cross-verified cancer risk barcode (**Fig.4g**). Although the most cases are distinctly classified into blue (~0) and red (~1) blocks, certain cases (e.g., C, 4, 11, and 2) exhibit unstable color blocks due to FP/FN predictions during cross-validation. Fortunately, the averaged barcode at the bottom of each model successfully reduces false cases, resulting in a more accurate probability estimation. This demonstrates the self-correction ability of the multi-modal cross-validation method. (**Fig. S33** and **S34**).

To simulate real diagnostic variability, analysis result from the compression method with full-data DNN predictions are compared in **Fig.4h**. The compression method's diagnostic data distribution was unordered, with significant overlap between positive and negative cases. By defining a 50% probability as the identifying threshold and cases within the 50 ± 20% range as suspected cases, the DNN-predicted data contains only 15% suspected cases — nearly three times fewer than those identified using the data compression method (**Fig.4i**). This substantially enhances diagnostic reliability by better differentiated positive and negative cases. Most notably, the prediction accuracy improved dramatically from 85% to 100% when using the DNN (**Figs. 4j** and **k**). Although the sample size was limited, the substantial improvement in lung cancer risk prediction achieved with DNN assistance not only boosts accuracy but also offers a clearer distinction between positive and negative cases for more reliable clinical diagnosis.

**Conclusion**

This work introduces a $Q$-switched sensing mechanism to address the longstanding challenge of refractometric biosensors with a high $Q$ factor but demands complicated illumination and detection setups. $Q$-switched sensing enabling rapid energy radiation variation to refractive index perturbation is realized in a unique 3D BIC metasurface with spatial-asymmetric structures. It provides an almost "in-situ" intensity response and excellent defect tolerance across visible to short-wave infrared spectrum. Utilizing wafer-scale aluminum-3D imprinting and direct stripping techniques, we have successfully mass-produced 8-inch 3D BIC metasurface chips. These chips demonstrate high $Q$ sensitivity up to $1.6 \times 10^4$ RIU$^{-1}$, covering illumination bandwidths from 2 nm to 25 nm. Therefore, it facilitates a spectrometer-free imaging with high resolution in vertical (~3.5 nm). Furthermore, miniaturized BIC biosensing system based on $Q$-switched sensing exhibits outstanding molecular detection performance to lung cancer-derived exosomes at concentrations as low as 129 aM. Combined with a DNN-assisted full-spectral analysis, nearly 100% prediction accuracy is achieved for cancer risk identification. The miniaturized BIC biosensing system underscore its potential for early-stage cancer diagnostics and commercialized platforms for clinical applications.

**References and Notes**

[1] A. M. Shrivastav, U. Cvelbar, & I. Abdulhalim. A comprehensive review on plasmonic-based biosensors used in viral diagnostics. *Commun. Biol.* **4**, 70 (2021).

[2] S. Sun, L. Wu, Z. Geng, P. P. Shum, X. Ma & J. Wang. Refractometric Imaging and Biodetection Empowered by Nanophotonics. *Laser Photonics Rev.* **17**, 2200814 (2023).

[3] I. Abdulhalim, M. Zourob & A. Lakhtakia. Surface plasmon resonance for biosensing: A mini-review. *Electromagnetics* **28**, 214–242 (2008).

[4] H. Altug, S. H. Oh, S. A. Maier, et al. Advances and applications of nanophotonic biosensors. *Nat. Nanotechnol.* **17**, 5–16 (2022).

[5] F. Li, J. Huang, C. Guan, et al. Affinity exploration of SARS-CoV-2 RBD variants to mAb-functionalized plasmonic metasurfaces for label-free immunoassay boosting. *ACS Nano* **17**, 3383–3393 (2023).

[6] H. Zhou, et al. Surface plasmons-phonons for mid-infrared hyperspectral imaging. *Sci. Adv.* **10**, eado3179 (2024).

[7] D. Rodrigo, et al. Mid-infrared plasmonic biosensing with graphene. *Science* **349**, 165–168 (2015).

[8] R. M. Kim, J. H. Huh, S. Yoo, et al. Enantioselective sensing by collective circular dichroism. *Nature* **612**, 470–476 (2022).

[9] Y. Shen, J. Zhou, T. Liu, et al. Plasmonic gold mushroom arrays with refractive index sensing figures of merit approaching the theoretical limit. *Nat. Commun.* **4**, 2381 (2013).

[10] A. Moreau, C. Ciracì, J. Mock, et al. Controlled-reflectance surfaces with film-coupled colloidal nanoantennas. *Nature* **492**, 86–89 (2012).

[11] C. Zhang, et al. Switching plasmonic nanogaps between classical and quantum regimes with supramolecular interactions. *Sci. Adv.* **8**, eabj9752 (2022).

[12] A. Xomalis, X. Zheng, R. Chikkaraddy, et al. Detecting mid-infrared light by molecular frequency upconversion in dual-wavelength nanoantennas. *Science* **374**, 1268-1271 (2021).

[13] M. Limonov, M. Rybin, A. Poddubny, et al. Fano resonances in photonics. *Nat. Photon.* **11**, 543–554 (2017).

[14] N. Li, T. D. Canady, Q. Huang, et al. Photonic resonator interferometric scattering microscopy. *Nat. Commun.* **12**, 1744 (2021).

[15] A. Leitis, et al. Angle-multiplexed all-dielectric metasurfaces for broadband molecular fingerprint retrieval. *Sci. Adv.* **5**, eaaw2871 (2019).

[16] A. Tittl, et al. Imaging-based molecular barcoding with pixelated dielectric metasurfaces. *Science* **360**, 1105–1109 (2018).

[17] A. Aigner, et al. Plasmonic bound states in the continuum to tailor light-matter coupling. *Sci. Adv.* **8**, eadd4816 (2022).

[18] J. Hu, F. Safir, K. Chang, et al. Rapid genetic screening with high quality factor metasurfaces. *Nat. Commun.* **14**, 4486 (2023).

[19] L. Kühner, L. Sortino, R. Berté, et al. Radial bound states in the continuum for polarization-invariant nanophotonics. *Nat. Commun.* **13**, 4992 (2022).

[20] K. Watanabe & M. Iwanaga, Nanogap enhancement of the refractometric sensitivity at quasi-bound states in the continuum in all-dielectric metasurfaces. *Nanophotonics* **12**, 99–109 (2023).

[21] Y. Liang, K. Koshelev, F. Zhang, H. Lin, S. Lin, J. Wu, B. Jia, and Y. Kivshar. Bound States in the Continuum in Anisotropic Plasmonic Metasurfaces. *Nano Lett.* **9**, 6351–6356 (2022)

[22] D. C. Marinica, A. G. Borisov & S. V. Shabanov. Bound states in the continuum in photonics. *Phys. Rev. Lett.* **100**, 183902 (2008).

[23] K. Koshelev, S. Lepeshov, M. Liu, A. Bogdanov & Y. Kivshar. Asymmetric metasurfaces with high-Q resonances governed by bound states in the continuum. *Phys. Rev. Lett.* **121**, 193903 (2018).

[24] T. Santiago-Cruz, et al. Resonant metasurfaces for generating complex quantum states. *Science* **377**, 991–995 (2022).

[25] Y. Jahani, E. R. Arvelo, F. Yesilkoy et al. Imaging-based spectrometer-less optofluidic biosensors based on dielectric metasurfaces for detecting extracellular vesicles. *Nat. Commun.* **12**, 3246 (2021).

[26] F. Yesilkoy, E. R. Arvelo, Y. Jahani, et al. Ultrasensitive hyperspectral imaging and biodetection enabled by dielectric metasurfaces. *Nat. Photonics* **13**, 390–396 (2019).

[27] S. Ansaryan, Y. C. Liu, X. Li, et al. High-throughput spatiotemporal monitoring of single-cell secretions via plasmonic microwell arrays. *Nat. Biomed. Eng* **7**, 943–958 (2023).

[28] E. Balaur, S. O' Toole, A. J. Spurling, et al. Colorimetric histology using plasmonically active microscope slides. *Nature* **598**, 65–71 (2021).

[29] J. Kühne, J. Wang, T. Weber, et al. Fabrication robustness in BIC metasurfaces. *Nanophotonics* **10**, 4305–4312 (2021).

[30] D. Dmitriy, R. Sergey, R. Yury & N. Igor. Light–matter interaction in the strong coupling regime: configurations, conditions, and applications. *Nanoscale* **10**, 3589 (2018).

[31] W. K. Kim, J. Tang, P. H. Kuo & S. F. Kuo. Implementation and phase detection of dielectric-grating-coupled surface plasmon resonance sensor for backside incident light. *Opt. Express* **27**, 3867–3872 (2019).

[32] C. F. Doiron, I. Brener & A. Cerjan. Realizing symmetry-guaranteed pairs of bound states in the

continuum in metasurfaces. *Nat. Commun.* **13**, 7534 (2022).

[33] W. Wang, Y. K. Srivastava, T. C. Tan, et al. Brillouin zone folding driven bound states in the continuum. *Nat. Commun.* **14**, 2811 (2023).

[34] M., Cotrufo, A., Cordaro, D. L. Sounas, et al. Passive bias-free non-reciprocal metasurfaces based on thermally nonlinear quasi-bound states in the continuum. *Nat. Photon.* **18**, 81–90 (2024).

[35] L. Wen, R. Xu, Y. Mi, et al. Multiple nanostructures based on anodized aluminium oxide templates. *Nat. Nanotechnol.* **12**, 244–250 (2017).

[36] Z. Wang, J. Sun, J. Li, et al. Customizing 2.5D out-of-plane architectures for robust plasmonic bound-states-in-the-continuum metasurfaces. *Adv. Sci.* **10**, 2206236 (2023).

[37] X. Sun, J. Sun, Z. Wang, et al. Manipulating dual bound states in the continuum for efficient spatial light modulator. *Nano Lett.* **22**, 9982–9989 (2022).

[38] T. Weber, L. Kühner, L. Sortino et al. Intrinsic strong light-matter coupling with self-hybridized bound states in the continuum in van der Waals metasurfaces. *Nat. Mater.* **22**, 970–976 (2023).

[39] C. Yang, W. Chen, X. Kong, D. Wang, J. Chen & C. W. Qiu. Can weak chirality induce strong coupling between resonant states? *Phys. Rev. Lett.* **128**, 146102 (2022).

[40] S. I. Azzam, V. M. Shalaev, A. Boltasseva & A. V. Kildishev. Formation of bound states in the continuum in hybrid plasmonic-photonic systems. *Phys. Rev. Lett.* **121**, 253901 (2018).

[41] V. Dolia, H. B. Balch, S. Dagli, et al. Very-large-scale-integrated high-quality factor nanoantenna pixels. *Nat. Nanotechnol*. 1748-3395 (2024).

[42] Chen, W., Kaya Özdemir, Ş., Zhao, G. et al. Exceptional points enhance sensing in an optical microcavity. *Nature* **548**, 192–196 (2017).

[43] Wenbo, M., Zhoutian, F., Fu, L., and Lan, Y. Exceptional–point–enhanced phase sensing. *Sci. Adv.* **10**, eadl5037 (2024).

[44] Bolognesi, M., Prosa, M., Toerker, M., Lopez Sanchez, L., et al. A Fully Integrated Miniaturized Optical Biosensor for Fast and Multiplexing Plasmonic Detection of High- and Low-Molecular-Weight Analytes. *Adv. Mater.* **35**, 2208719 (2023).

[45] O. Myriam, S. Esther et al. Point-of-care detection of extracellular vesicles: Sensitivity optimization and multiple-target detection. *Biosensors and Bioelectronics.* **87**, 38-45 (2017).

[46] H. Liang, X. Wang, F. Li, et al. Label-free plasmonic metasensing of PSA and exosomes in serum for rapid high-sensitivity diagnosis of early prostate cancer. *Biosensors and Bioelectronics*. **235**, 0956-5663 (2023).

[47] L. Xu, R. Chopdat, et al. Development of a simple, sensitive and selective colorimetric aptasensor for the detection of cancer-derived exosomes. *Biosensors and Bioelectronics* **169**, 112576 (2020).

[48] Z. Wang, S. Zong, Y. Wang, et al. Screening and multiple detection of cancer exosomes using SERS-based method. *Nanoscale* **10**, 9053 (2018).

[49] J. Chen, Y. Xu, Y. Lu and W. Xing. Isolation and visible detection of tumor-derived exosomes from plasma. *Analytical Chemistry* **90**, 14207-14215 (2018).

[50] D. Jin, F. Yang, Y. Zhang, et al. ExoAPP: Exosome-oriented, aptamer nanoprobe-enabled surface proteins profiling and detection. *Analytical Chemistry* **90**, 14402−14411 (2018).

[51] P. Zhang, X. Zhou, M. He, et al. Ultrasensitive detection of circulating exosomes with a 3D-nanopatterned microfluidic chip. *Nat Biomed Eng* **3**, 438–451 (2019).

[52] R. Huang, L. He, Y. Xia, et al. A Sensitive aptasensor based on a hemin/G-quadruplex assisted signal amplification strategy for electrochemical detection of gastric cancer exosomes. *Small* **15**, 1900735 (2019).

[53] H. Xiong, Z. Huang, Q. Lin, et al. Surface plasmon coupling electrochemiluminescence immunosensor based on polymer dots and AuNPs for ultrasensitive detection of pancreatic cancer exosomes. *Analytical Chemistry* **94**, 837-846 (2022).

[54] H. Rongsheng, L. Shaowei, L. Fajun, et al. Exploring aptamer-based metasurfaces for label-free plasmonic biosensing of breast tumor-derived exosomes. *Adv. Opt. Mat.* 2401180 (2024).

[55] J. Park, J. S. Park, C. H. Huang, et al. An integrated magneto-electrochemical device for the rapid profiling of tumour extracellular vesicles from blood plasma. *Nat Biomed Eng* **5**, 678–689 (2021).

[56] C. Liu, J. Zhao, F. Tian, et al. Low-cost thermophoretic profiling of extracellular-vesicle surface proteins for the early detection and classification of cancers. *Nat Biomed Eng* **3**, 183–193 (2019).

[57] X. Wu et al. Exosome-templated nanoplasmonics for multiparametric molecular profiling. *Sci. Adv.* **6**, eaba2556 (2020).

**Acknowledgments:** The authors thank the facility support and technical assistance from the Westlake Centre for Micro/Nano Fabrication, the Instrumentation and Service Centre for Physical Sciences (ISCPS), and the Instrumentation and Service Centre for Molecular Sciences (ISCMS) at Westlake University.

**Funding:**

Natural Science Foundation of China (Grants No. 52373238, 52003225, 62175205, and U2130112)

Research Centre for Industries of the Future at Westlake University (RCIF, Grant No. WU2022C024)

Special Support Plan for Photoelectric Chips Research at Westlake University (Grant No. 10300000H062201)

Key Project of Westlake Institute for Optoelectronics (Grant No. 2023GD005) and Westlake Education Foundation

The Youth Talent Support Program of Fujian Province (Eyas Plan of Fujian Province) [2022]

**Author contributions:** J.C.S and L.Y.W conceived the idea. J.F.Z and L.Y.W supervised the project. M.Q provided helpful discussions. J.C.S implemented the theoretical analysis, simulation design, fabrication, optical measurement, and deep learning algorithms. J.C.S and J.H built the optical measurement setup. X.D.W and J.C.S implemented the miniaturized biosensing system. F.J.L and S.W.L provided the bio-samples. F.J.L and X.D.W implemented the microfluidic biosensing test and data collection. J.C.S, F.J.L, X.D.W and L.Y.W performed the data process and analysis. J.C.S and L.Y.W wrote the manuscript with input from J.F.Z, F.J.L, and M.Q. All authors contributed to the manuscript and approved the final version.

**Competing interests:** The authors declare that they have no competing interests.

**Data and materials availability:** All data are available in the main text or the supplementary

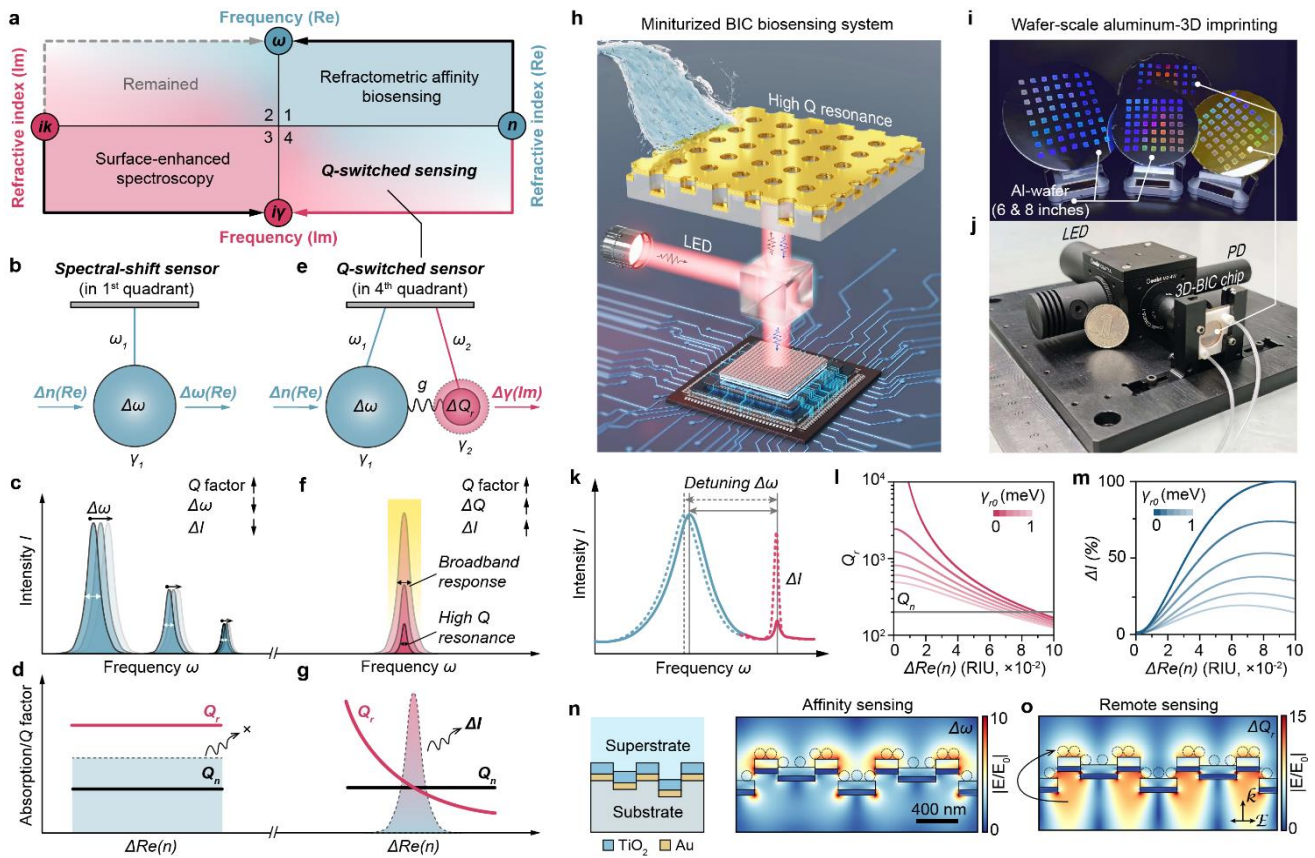

**Fig. 1. Schematic of *Q*-switched sensing mechanism and fully-integrated biosensing system enabled by 3D-BIC metasurface. a** Phase diagram of different photonic biosensing mechanisms: refractometric affinity biosensing at 1st quadrant, surface-enhanced spectroscopy biosensing at 3rd, *Q*-switched sensing proposed in this work at 4th quadrant and the sensing mechanism at 2nd quadrant is remained to be explored. **b-d** Schematic view of single oscillator sensing mechanism with real part of frequency change (*Δw*) in response to real part of refractive index and its corresponding spectral behavior with higher but constant *Q* factor. **e-g** Schematic view of two coupled oscillators sensing mechanism with imaginary part of frequency (*Δγ*) in response to real part of refractive index and its corresponding spectral behavior with higher but dynamic *Q* factor. **h** Schematic of miniaturized *Q*-switched biosensing system with 3D-BIC metasurface. **i** Wafer-scale fabrication of 3D-BIC metasurface by aluminum-based 3D nanoimprinting method. **j** Picture of real products of miniaturized 3D-BIC biosensing system. **k** Spectral response of strong coupling system to *ΔRe(n)* with *Q*-switched sensing. **l** and **m** are analytical results of radiative *Q* factor, $Q_r$ and peak intensity change *ΔI* from *Q*-switched sensing equation derived from strong coupling system. **n** Schematic of 3D-BIC metasurface and its electric field profile of affinity sensing mode with *Δw*. **o** Electric field profile of *Q*-switched sensing mode with $ΔQ_r$.

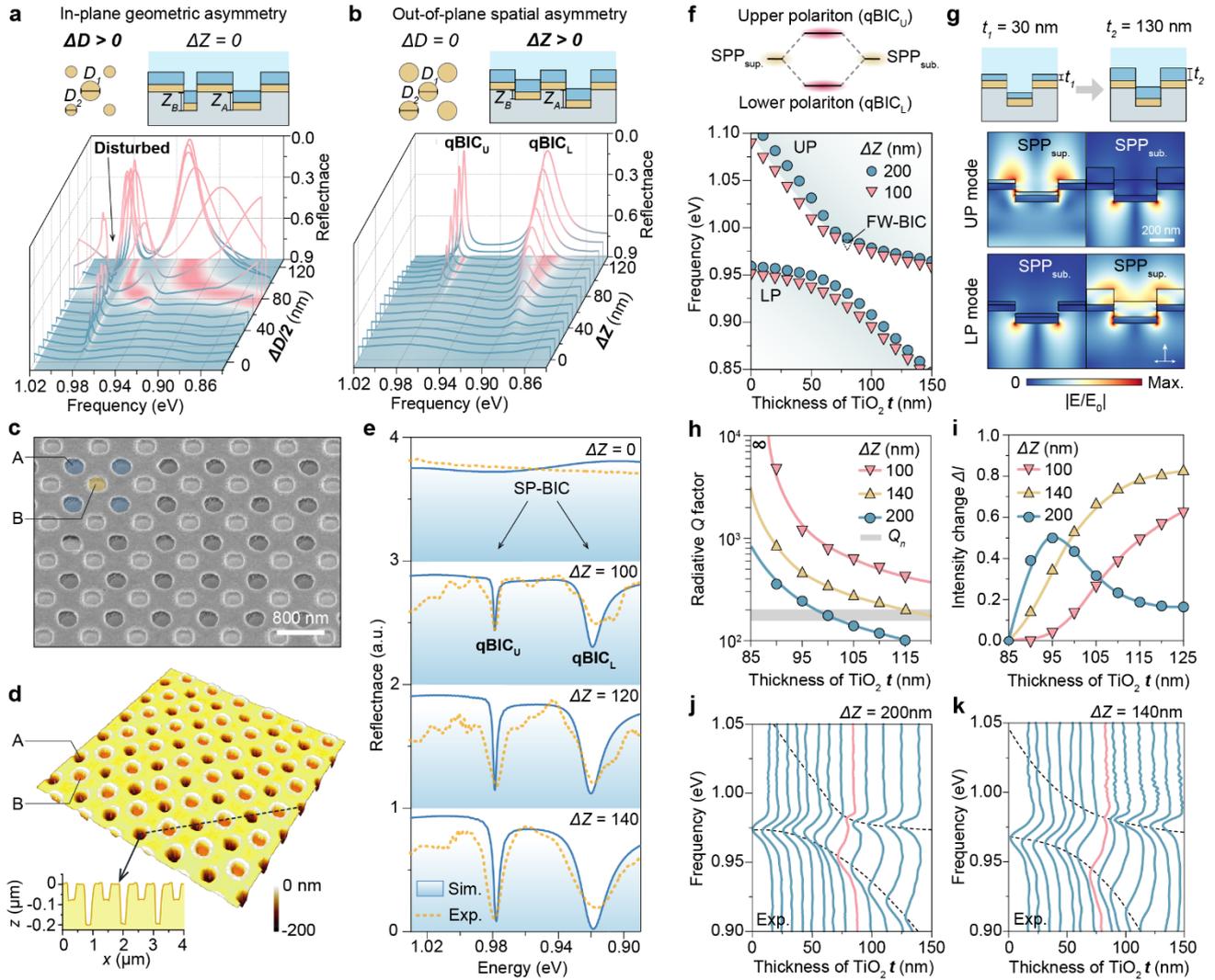

**Fig. 2. Structure parameter optimization of 3D-BIC metasurface in strong coupling system. a,b** Simulation comparison between introducing in-plane asymmetry and out-of-plane asymmetry for manipulating qBIC resonances in strong coupling system. It can be observed that conventional in-plane asymmetry will causing mode-crosstalk between multiple resonances, while the out-of-plane 3D-BIC configuration can perfectly avoid such crosstalk, providing robust qBIC manipulation. **c,d** SEM and AFM characterization of 3D-BIC metasurface morphology. **e** Measured and simulated spectra of qBIC resonances with different height difference $\Delta Z$ ($Z_A$-$Z_B$). **f** Simulated strong coupled mode resonances with increasing of $TiO_2$ thickness. A smaller $\Delta Z$ induces more constrained radiative loss will cause the appearance of FW-BIC with larger Rabi splitting. **g** Simulated mode conversion between SPP mode at superstrate ($SPP_{sup}$) and SPP mode at substrate ($SPP_{sub}$) with different $TiO_2$ thickness. **h,i** Simulated $Q$-switched process and peak intensity change with the increase of $TiO_2$ thickness. **j,k** Measured spectra show typical strong coupling behavior where the Rabi splitting increases with smaller $\Delta Z$.

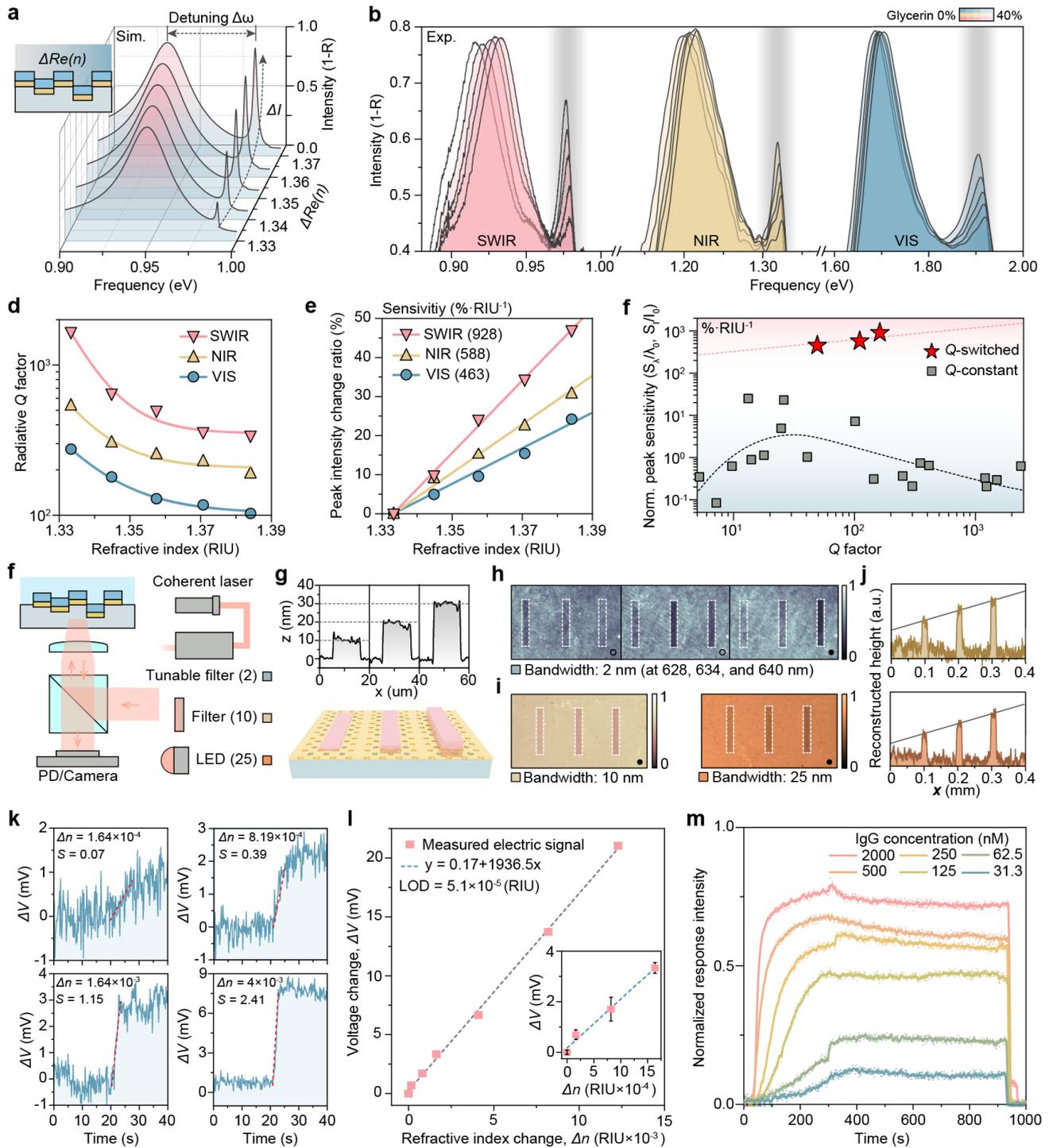

**Fig. 3. Bulk sensing performance of *Q*-switched sensor and its hyperspectral image characterization for surface sensing. a** Simulated *Q*-switched bulk sensing performance with *ΔRe(n)*, which shows significantly growing of peak intensity of upper branch with small frequency detuning. **b** Experimental achieved *Q*-switched bulk sensing at visible (VIS), near-infrared (NIR) and short-wave infrared (SWIR) with the increase of glycerin concentration in DI water. **c** Mode distributions of lower branch with frequency shift (*Δw*) in response to *ΔRe(n)* which obeys typical affinity sensing profile and upper branch with switched *Q* factor (*ΔQ_r*) whose major mode is concentrated in the substrate, while can remotely sensing the *ΔRe(n)* happens in superstrate. This is the primary difference of mode distribution between *Q*-switched sensing and frequency-shift sensing. **d** Radiative *Q* factor response

to $\Delta Re(n)$ derived from experimental data. **e** Peak intensity response and sensitivity to $\Delta Re(n)$ derived from experimental data. **f** References summary and comparison with normalized peak response sensitivity between conventional frequency-shift ($Q$-constant) sensors and our $Q$-switched (frequency-constant) sensor. It can be seen that the frequency sensitivity inevitably decreases with the increase of $Q$ factor for frequency-shift sensors, while the $Q$-switched sensor could exhibit better performance with higher $Q$ factor and has peak sensitivity much higher than the $Q$-constant sensors. **g** Schematic of hyperspectral image system. **h** Microscale $SiO_2$ bars with thickness of 10, 20 and 30 nm are deposited on the 3D-BIC metasurface. **i** Reconstructed images for $SiO_2$ bars illuminated by narrowband tunable light source. **j** Reconstructed images for $SiO_2$ bars illuminated by broadband light sources. **k** Reconstructed height of $SiO_2$ bars from broadband illumination, which confirms the compatibility to various bandwidths light sources of $Q$-switched sensor. **k** Dynamic curves of the integrated system response to different $\Delta Re(n)$. **l** Fitted bulk sensing data for $\Delta Re(n)$ from different sucrose concentration. **m** Dynamic fitted curves of the interaction between ProA on the surface of the 3D-BIC metasurface sensor with different concentrations of IgG.

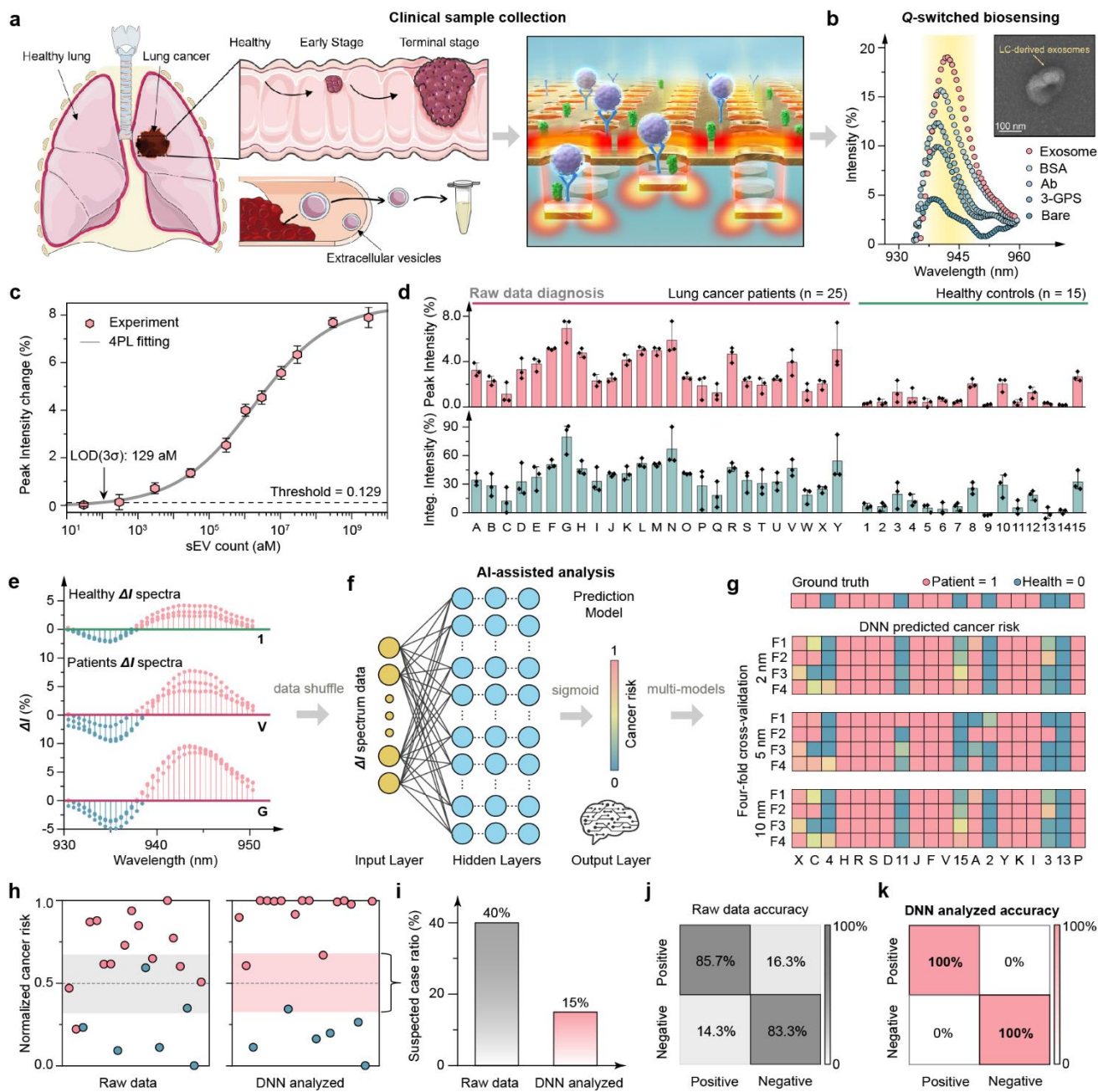

**Fig. 4.** *Q*-switched sensor applied in clinical lung cancer diagnosis with AI assistance. **a** Schematic of clinical lung cancer diagnosis utilizing 3D-BIC metasurface. **b** *Q*-switched peak intensity response during bio-functional steps. The inset shows representative TEM images of purified LC-derived exosomes with sizes ranging from ~160 nm. **c** *Q*-switched peak intensity response as a function of exosome concentrations, fitted by the four-parameter logistic (4PL) equation. **d** List of peak intensity changes and integrated intensity change induced by serum samples from cancer patients and healthy controls. **e** Intensity spectral data collected in triplicate for each healthy control and patient as input data for the DNN model. **f** DNN model applied for improving cancer diagnosis, using spectral intensity change data from 2, 5, and 10 nm bandwidths as input matrix and outputting a predicted cancer risk from 0 to 1. **g** Training process of DNN model with 4-fold cross-validation, compared to the ground truth color bar code. **h** Normalized cancer risk distribution comparison between raw data and DNN analyzed results. **i** Comparison of suspected cases number between raw data and DNN analyzed results

within a range of 50 ± 20% risk value. **m,n** Confusion matrix of classification from raw data and DNN analyzed results.